\documentclass[aps,prl,twocolumn,groupedaddress,showpacs]{revtex4}
\usepackage{natbib}
\usepackage{bm}
\usepackage{graphicx}
\begin{document}

\title{Detection of primordial non-Gaussianity ($f_{\rm NL}$) in the WMAP 3-year
data at above 99.5\% confidence}

\author{
Amit P. S. Yadav$^1$ and Benjamin D. Wandelt$^{1,2}$}
\affiliation{$^1$Department of Astronomy, University of Illinois at 
Urbana-Champaign, 1002 W.~Green Street, Urbana, IL 61801}

\affiliation{$^2$Department of Physics,  University of Illinois at Urbana-Champaign, 1110 W.~Green Street, Urbana, IL 61801}

\begin{abstract}
We present evidence for the detection of primordial non-Gaussianity of the local
type ($f_{\rm NL}$), using the temperature information of the Cosmic Microwave
Background (CMB) from the WMAP 3-year data. We employ the bispectrum estimator
of non-Gaussianity described in~\cite{linear_term07} which allows us to analyze
the entirety of the WMAP data without an arbitrary cut-off in angular scale.
Using the combined information from WMAP's two main science channels up to
$\ell_{max}=750$ and the conservative Kp0 foreground mask we find  $27  < \mathbf {f_{\rm NL}} <
147$ at $95 \%$ C.L., with a central value of $f_{\rm NL}=87$. This corresponds
to a rejection of $f_{\rm NL}=0$ at more than 99.5\% significance. We find that this
detection is robust to variations in $l_{max}$, frequency and masks, and that 
no known foreground, instrument systematic, or secondary anisotropy explains our
signal while passing our suite of tests.  We explore the
impact of several analysis choices on the stated significance and find 2.5
$\sigma$ for the most conservative view.
We conclude that the WMAP 3-year data disfavors canonical single field slow-roll
inflation.
\end{abstract} 
\maketitle
 
It is now widely accepted that tests of primordial non-Gaussianity, parameterized by the non-linearity parameter $f_{\rm NL}$, promise to be a unique probe of the early Universe~\cite{KS2001} beyond the two-point statistics. Although the non-Gaussianity from the simplest inflation models is very
small, $f_{\rm NL}\sim 0.01-1$ ~\citep{Salopek_Bond90,Gangui_etal94,small_nonG},
there is a  very large  class  of  more  general models, e.g., models with
multiple scalar fields, features in inflation potential, non-adiabatic
fluctuations,    non-canonical   kinetic   terms,    deviations   from
the Bunch-Davies vacuum, among others, that predict  substantially   higher
level of primordial non-Gaussianity (see \cite{BKMR_04} for a review and detailed references). 

Recent calculations of the perturbations arising in the ekpyrotic or cyclic
cosmological
scenarios~\cite{Steinhardt_Turok_02a} have
concluded that these scenarios can predict $f_{\rm NL}$ much larger than single
field slow-roll
inflation~\cite{Creminelli_Senatore07}. Detailed calculations in these models are fraught with difficulties connected to matching the perturbations through the cosmological singularity at the bounce. However, the current calculations suggests that primordial non-Gaussianity of the $f_{\rm NL}$ type could be a powerful discriminant between ekpyrotic models and standard slow-roll inflation. As such, the search for primordial non-Gaussianity is complementary to the search for the inflationary gravitational wave background. We will argue in this letter that the WMAP 3-year data already distinguishes $f_{\rm NL}=100$ from $f_{\rm NL}\sim 0$ at a statistically significant level.

Primordial non-Gaussianity can be described in terms of the 3-point
correlation function of Bardeen's curvature perturbations, $\Phi(k)$, in
Fourier space: 
\begin{eqnarray}
\langle \Phi(\mathbf{k_1})(\mathbf{k_2})(\mathbf{k_3})\rangle = (2\pi)^3\delta^3(\mathbf{k_1} + \mathbf{k_2} + \mathbf{k_3})F(k_1, k_2, k_3).
\end{eqnarray}
Depending on the shape of the 3-point function, i.e., $F(k_1, k_2,
k_3)$, non-Gaussianity can be broadly classified into two
classes~\citep{Babich_etal_04}. 
First, the local, ``squeezed,'' non-Gaussianity where
$F(k_1, k_2, k_3)$ is large for the configurations in which $k_1 \ll k_2,
k_3$. Second, the non-local, ``equilateral,'' non-Gaussianity where $F(k_1, k_2, k_3)$ is
large for the configuration when $k_1 \sim k_2 \sim k_3$. 

The local form arises from a non-linear relation between
inflaton and curvature
perturbations~\citep{Salopek_Bond90,Gangui_etal94},
curvaton models~\citep{Linde_Mukhanov1997}, 
or the ekpyrotic
models~\citep{Creminelli_Senatore07}. 
The equilateral form arises from non-canonical kinetic terms such 
as the Dirac-Born-Infeld (DBI)
action~\citep{Alishahiha_etal04}, the ghost
condensation~\citep{Arkani_et_04}, or any other single-field models in
which the scalar field acquires a low speed of sound
~\citep{Chen_etal07}.
While we focus on the local form in this letter, it is straightforward to
repeat our analysis for the equilateral form.

The local form of non-Gaussianity may be parametrized in real space
as~\citep{Gangui_etal94,verde00,KS2001}: 
\begin{equation}
\label{eqn:phiNG}
\Phi(\mathbf{r}) = \Phi_L(\mathbf{r}) + f_{\rm NL} \left( \Phi_L^2(\mathbf{r}) - \langle \Phi_L^2(\mathbf{r}) \rangle
\right)
\end{equation}
where $f_{\rm NL}$ characterizes the amplitude of primordial
non-Gaussianity. Note that the Newtonian potential has the opposite sign of
Bardeen's curvature perturbation, $\Phi$. 

The first fast bispectrum based $f_{\rm NL}$ estimator using temperature
anisotropies alone was introduced in~\cite{KSW05} (the KSW estimator). The idea
of adding a linear term to reduce excess variance due to 
noise inhomogeneity was introduced in~\cite{creminelli_wmap1}.  Applied to a
combination of the Q, V and W channels of the WMAP
3-year data up to $\ell_{max}\sim 400$ this estimator has yielded the tightest
constraint on 
$f_{\rm NL}$ so far: $-36 < f_{\rm NL} <100$ (2$\sigma$)~\cite{creminelli_wmap2}.
This estimator was generalized to utilize both the temperature and
E-polarization information in~\cite{linear_term07}, where we pointed out
that the linear term had been incorrectly implemented in Eq. 30 of
~\cite{creminelli_wmap1}. The corrected estimator enables us to analyze the entire WMAP data without suffering from a blow-up in the variance at high $\ell$.

\begin{figure}[!t]
\includegraphics[width=3.5in]{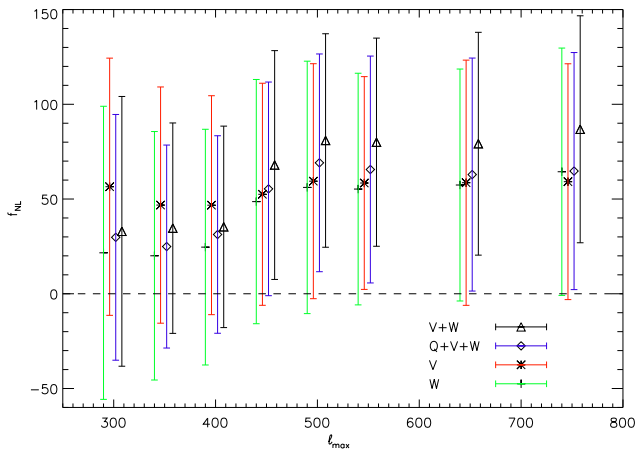}
\caption{We show the measured value of the non-linear coupling parameter $f_{\rm NL}$ using WMAP 3-year maps, and the corresponding $95 \%$ error bars derived from the Gaussian simulations. For this analysis the WMAP Kp0 mask was used. The analysis is done for 4 combinations of the frequency channels: coadded Q+V+W, coadded V+W, V, and W.}
\label{mean_fnl}
\end{figure}

{\it Our analysis.} We assume a standard Lambda CDM cosmology with following cosmological parameters: 
$\Omega_b = 0.042$, $\Omega_{cdm} = 0.239$, $\Omega_L = 0.719$, 
$h = 0.73$,
$\tau = 0.09$, and $n_s=1$. We will discuss the effect of varying these fiducial
parameters below.

We used the generalized bispectrum estimator of primordial non-Gaussianity of local type described in~\cite{linear_term07}. The generalized estimator is given by 
\begin{equation}
 \hat{f}_{NL} =
\frac{\hat{S}_{prim}+\hat{S}^{linear}_{prim}}{N},
\end{equation}
where $N$ is the normalization factor and $\hat{S}_{prim}$ and $\hat{S}^{linear}_{prim}$ are the so called trilinear and linear term of the estimator respectively. The trilinear term captures the bispectrum information about $f_{\rm NL}$ while the linear term has vanishing expectation and is designed to reduce the scatter in the trilinear term induced by the foreground mask and WMAP's anisotropic scan strategy.

Although our estimator~\cite{linear_term07} can utilize both the temperature and
E-polarization information of the CMB to constrain primordial non-Gaussianity,
we have used only temperature information of the WMAP 3-year data. For the
analysis we used various combinations of 8 channels of WMAP
3-year raw data: Q1, Q2, V1, V2, W1, W2, W3, and W4. For all the simulations we
used the WMAP 3-year maps in HEALPix format  with $N_{pix}=3145728$ pixels.
We focused on the V and W bands, which are the main WMAP CMB science channels suffering least
from foreground contamination. We also applied our
estimator to Q and Q+V+W to assess sensitivity to foregrounds. 

We performed Monte Carlo simulations to assess the statistical significance and errors of our $f_{\rm NL}$ estimates. For example for the Q+V+W coadded simulated map, we first simulated 8 Gaussian maps using the noise and beam properties of the corresponding 8 channels. Then a single map was obtained by pixelwise averaging of these 8 maps. The same procedure was followed to obtain simulated coadded maps of the other channel combinations. The $S_{AB}$  and $S_{BB}$ weight maps for the linear estimator \citep{creminelli_wmap1} were obtained using 800 Monte Carlo simulations that include the WMAP noise and foreground masks.   

\begin{table}[t]
\begin{center}
\begin{tabular}{|c|cccc|c|cccc|}
\hline
\cline{2-10} $\mathbf {\ell_{\rm max}}$ & \multicolumn{4}{|c|}{VW} &Q &\multicolumn{4}{|c|}{QVW} \\
\cline{2-10}    
 &Kp12 & Kp2 & Kp0 & Kp0+&Kp0 & Kp12 & Kp2 & Kp0 & Kp0+\\
\hline
350 & -1290  &  -27  &  35   &    19  &  1  &-2384 &  -75  &  25    &      8    \\ 
450 &  -1425  &   -16  &  68   &   65 & -6  &-2792  &  -80  &  55    &      65    \\
550  &  -1510  &  -13  & 80   &    84&-11 &-3136  &  -94  &  66    &      80  \\
650  & -1560 &   -22 &  79   &      81& -14 &-3307  &  -94   &  63    &      77  \\
750 & -1575  &   -23  &  87   &       87 &-20  &-3368  &  -108 &  65    &      78   \\
\hline
$750^\star$  &-1105$\pm^{19}_{19}$& -42$\pm^{5}_{5}$&  -6$\pm^{4}_{4}$ &  -0.3$\pm^{4}_{4}$  & &  & & -13$\pm^5_5$ & 1$\pm^6_6$ \\

\hline


\end{tabular}\caption{Non-linear coupling parameter $f_{\rm NL}$ using the V+W, Q, and Q+V+W WMAP 3-year raw maps, as a function of maximum multipole used in the analysis $\ell_{max}$ and mask Kp12, Kp2, Kp0, and Kp0+ (corresponding $f_{sky}$ is stated in the text and the masks are shown in Fig 2).  The last row (750$^\ast$) shows the mean $f_{\rm NL}$ estimated from Gaussian simulations including the WMAP foreground model. Foreground contamination biases $f_{\rm NL}$ negatively by similar amounts in both the data and the model.}
\label{QVW_fnl_mask}
\end{center}
\end{table}

Figure~\ref{mean_fnl} shows the measured value of the non-linear coupling parameter $f_{\rm NL}$  for 4 combinations of coadded frequency channels (Q+V+W, V+W, V, and W) as a function of maximum multipole $\ell_{max}$ used in the analysis. All the analyses in this figure use the Kp0 mask. The figure shows the $95\%$ C.L. error bars derived from Monte Carlo simulations.

For the coadded V+W map there is evidence of primordial non-Gaussianity
at more than $95\%$ C.L. for all $\ell_{max} > 450$.  For the coadded Q+V+W map
there is a detection of primordial non-Gaussianity at more than $95\%$ C.L. for
all $\ell_{max} > 500$. Residual suboptimality of our
estimator results in a larger error bar for the Q+V+W combination compared to the
V+W combination.  

Using the coadded V+W (the least foreground contaminated) channel with $\ell_{max}=750$, 
we find
\begin{equation}
27  < \mathbf {f_{\rm NL}} < 147 \;\;\;  \;\;\;({\rm at}\;\; 95 \% \;\;{\rm C.L.}).
\end{equation} This rules out the null hypothesis of Gaussian primordial
perturbations at 99.5\% significance. 

Our analysis provides the most information to date on the primordial non-Gaussianity of the local type. For the sake of comparison with the previous best result in the literature ($-36 < f_{\rm NL} <100$, for the coadded Q+V+W map at the $2\sigma$ level for $\ell_{max}\approx 400$
\citep{nong_wmap,creminelli_wmap2,wmap_2nd_spergel}), our constraints using the coadded Q+V+W map truncated at $\ell_{max}=400$ are: 
$ -20.84  < \mathbf {f_{\rm NL}} < 83.4 \;\;\;  \;({\rm at}\;\; 95 \% \;\;{\rm C.L.}).$
We may conclude that the additional information uncovered by the Yadav et al. estimator~\cite{linear_term07} at $\ell>400$ is important for our result.
As calculated by Creminelli et al.~\cite{Creminelli_estimators_nong} and
verified in simulation by~\cite{Liguori_Yadav_etal07}, there is a contribution
to the estimator variance due to non-zero $f_{\rm NL}$. This widens the confidence 
interval of the estimator by 3\%. It does not however modify
the significance of our rejection of the Gaussian null hypothesis.

{\it Interpretation.} A detection of non-Gaussianity has profound implications on our understanding of
the early Universe. We will now argue based on an extensive suite of null tests and
theoretical modeling that
our results are not due to any known systematic error,  foregrounds or secondary
anisotropy. 

Since our estimator is based on three-point correlations, any mis-specification
of the WMAP noise model would not bias our estimator, since Gaussian instrument
noise has a vanishing three-point function. Similarly, if the CMB were Gaussian,
asymmetric beams cannot create non-Gaussianity.  Beam far-side lobes can
produce a small level of smooth foreground contamination at high galactic latitude
\cite{wmap1_Barnes03} at $\ell\le 10$. This effect has been corrected in the 3-year
maps\cite{wmap2_Jarosik07}. Since our signal is not frequency dependent this is clearly not a dominant effect.
Even so, we checked for this or any other large scale anomaly (such as the axis
of evil) by deleting modes with $\ell\le20$ from our analysis. We find that our estimate
increases to $f_{\rm NL}=135\pm 96$ at (95 $\%$ C.L.), leaving the statistical
significance of our signal at a similar level.

To test for non-Gaussian time-domain systematics or non-Gaussian noise, 
we take difference between the pairs of yearly WMAP data. This creates 3 jackknife realizations of WMAP noise
maps for each detector including real instrument systematics.
Applying our estimator to these maps gave negligible $f_{\rm NL}$ ($\sim 1$) for all
three pairs of years, leading us to a conservative bound on the systematic error arising from such
effects of $\pm 2$. 

Seeing the same behavior in all channel combinations suggests that even if the
detection of non-Gaussianity were not primordial, its source would not be
frequency dependent and not associated with the main galactic foregrounds near the galactic plane. 
This is confirmed by repeating our analyses with several foreground  masks (WMAP Kp12 mask with $\mathbf f_{\rm sky}=94.2\%$, WMAP Kp2 mask with $\mathbf f_{\rm sky}=84.7\%$, WMAP Kp0 mask with $\mathbf f_{\rm sky}=76.8\%$, and a larger mask (Kp0+)  with $\mathbf f_{\rm sky}=64.3\%$, which was obtained by smoothing and thresholding the Kp0 mask) for computing the value of $f_{\rm NL}$ using the Q, Q+V+W and V+W maps. The results for these analyses are shown in Table~\ref{QVW_fnl_mask}. It is clear from these analyses that the Kp12 and Kp2 masks do not exclude galactic foregrounds at the required level. The Q+V+W combination is foreground contaminated even for Kp0 mask, as is also clear from Q band analysis.  However, for V+W, increasing the mask beyond Kp0 does not change the results significantly. 

\begin{figure}[!t]
\includegraphics[height=1.5 in,angle=90 ]{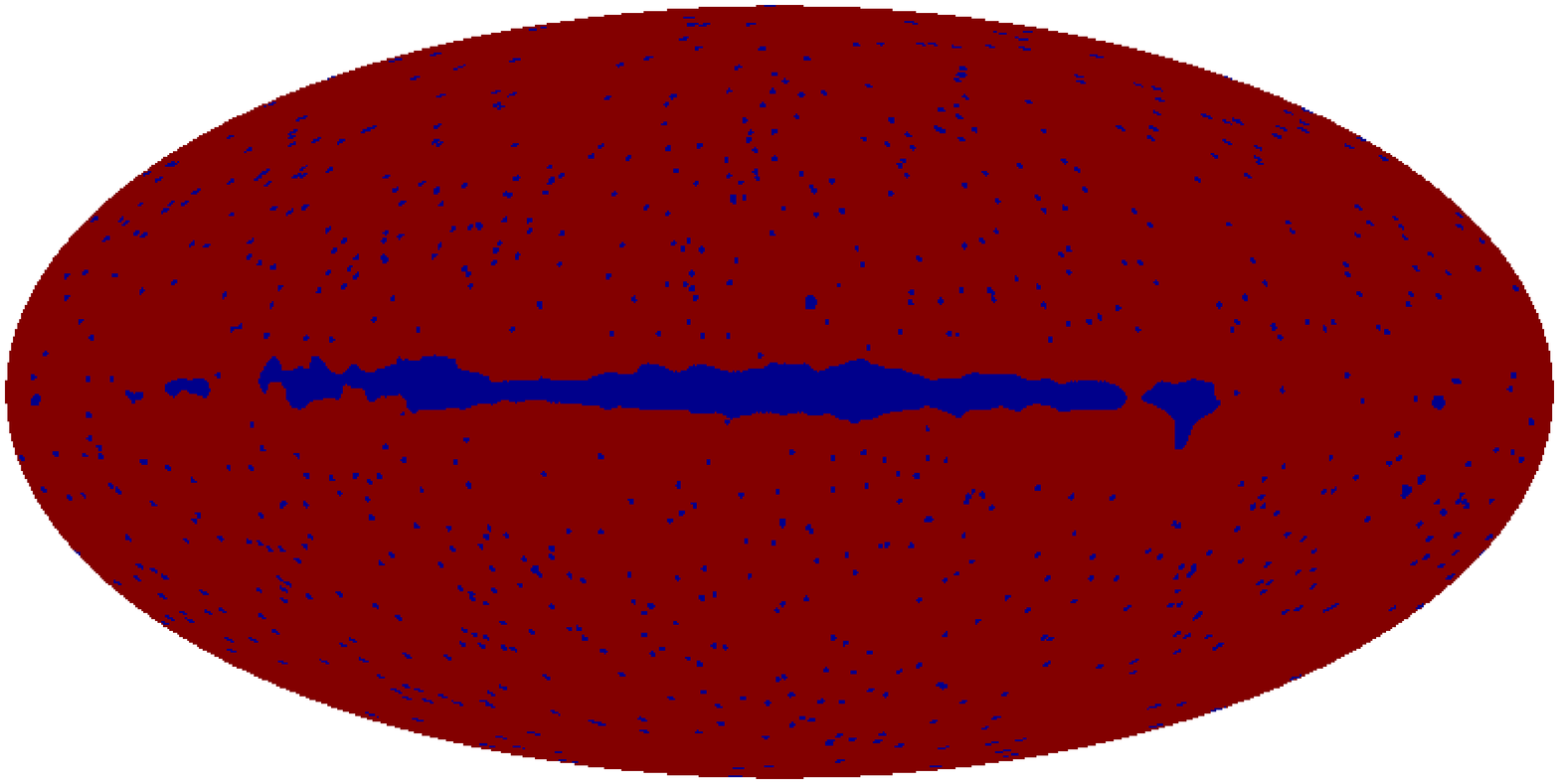}
\includegraphics[height=1.5 in,angle=90 ]{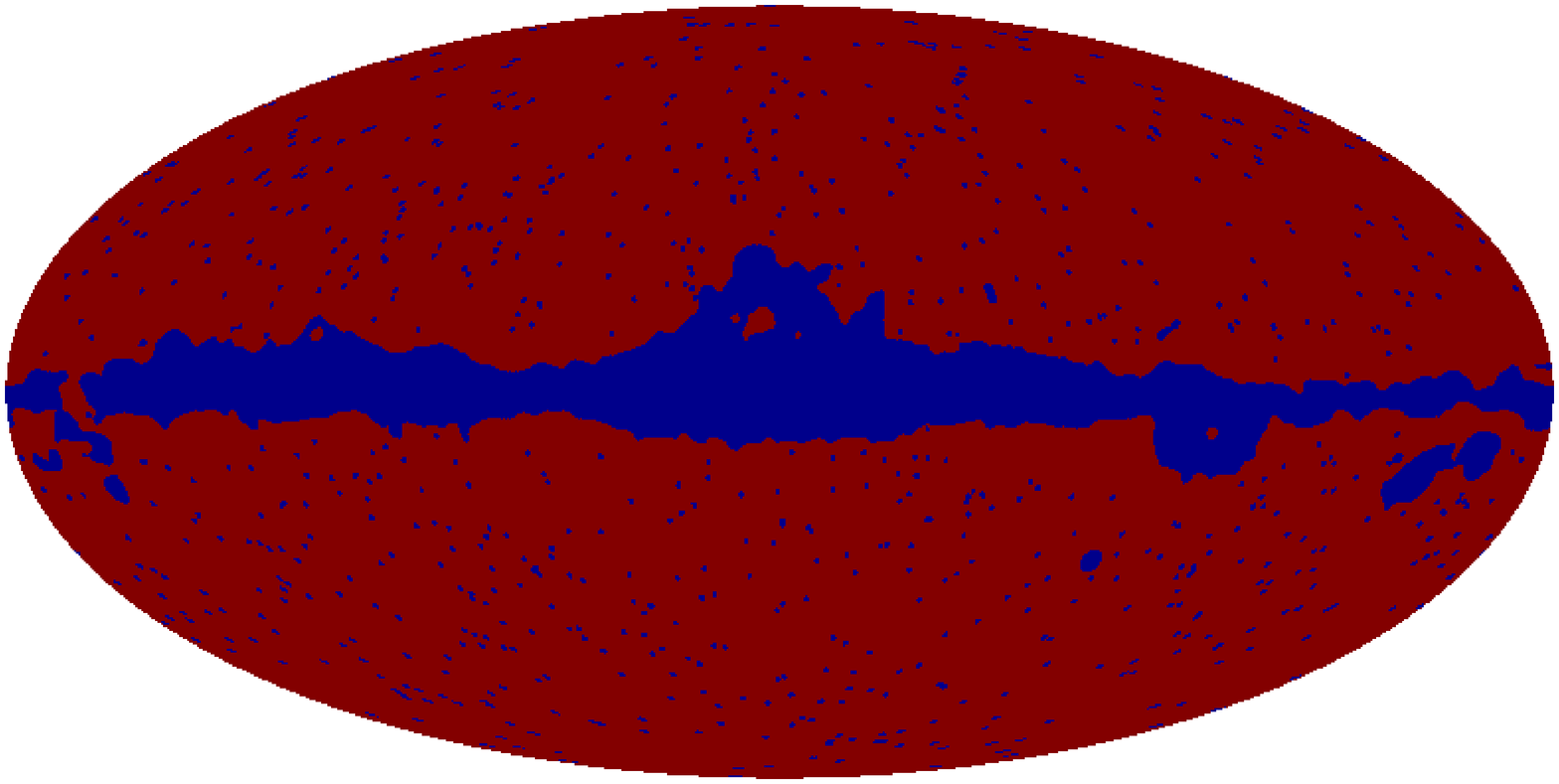}
\includegraphics[height=1.5 in,angle=90 ]{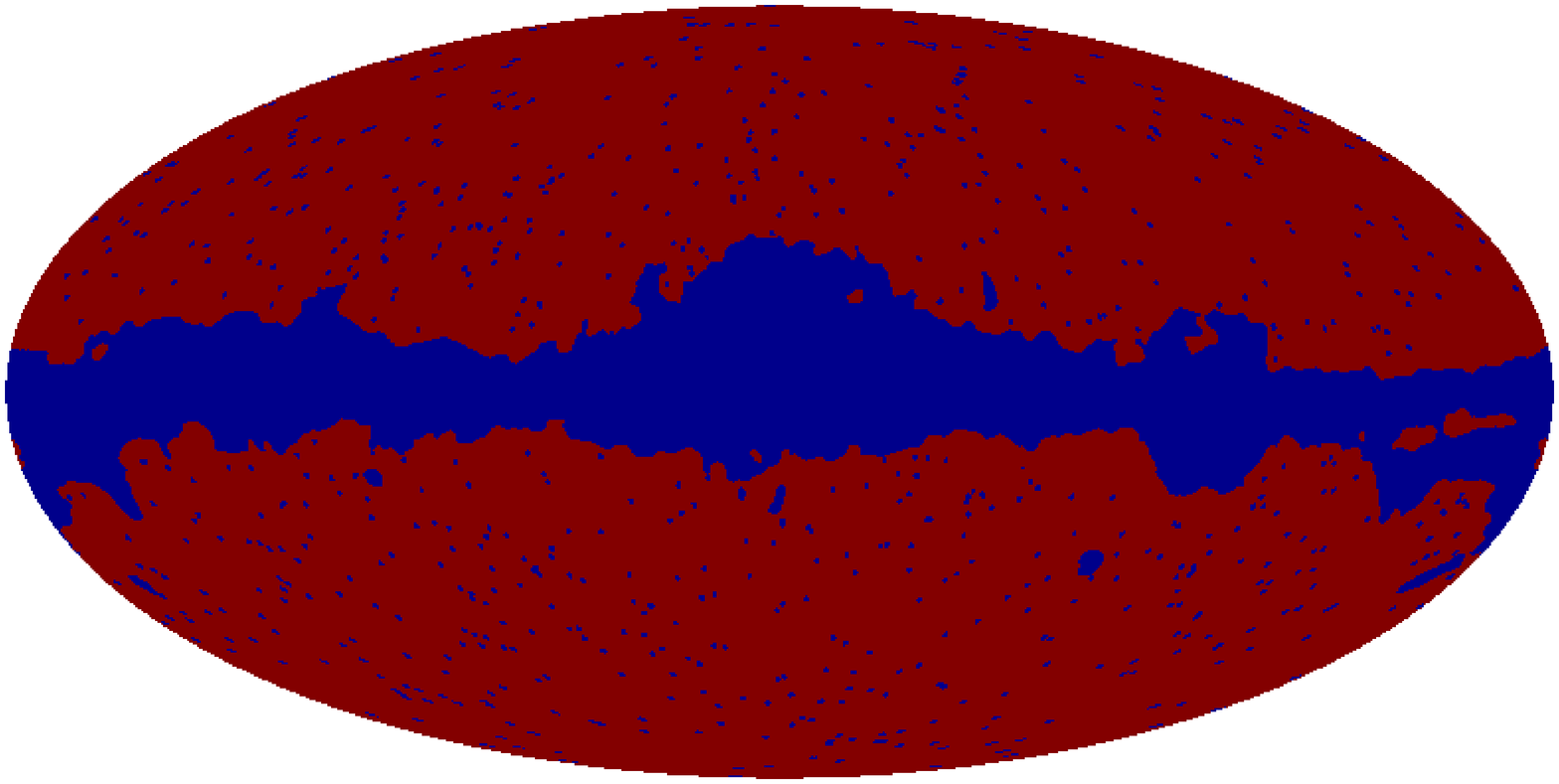}
\includegraphics[height=1.5 in,angle=90 ]{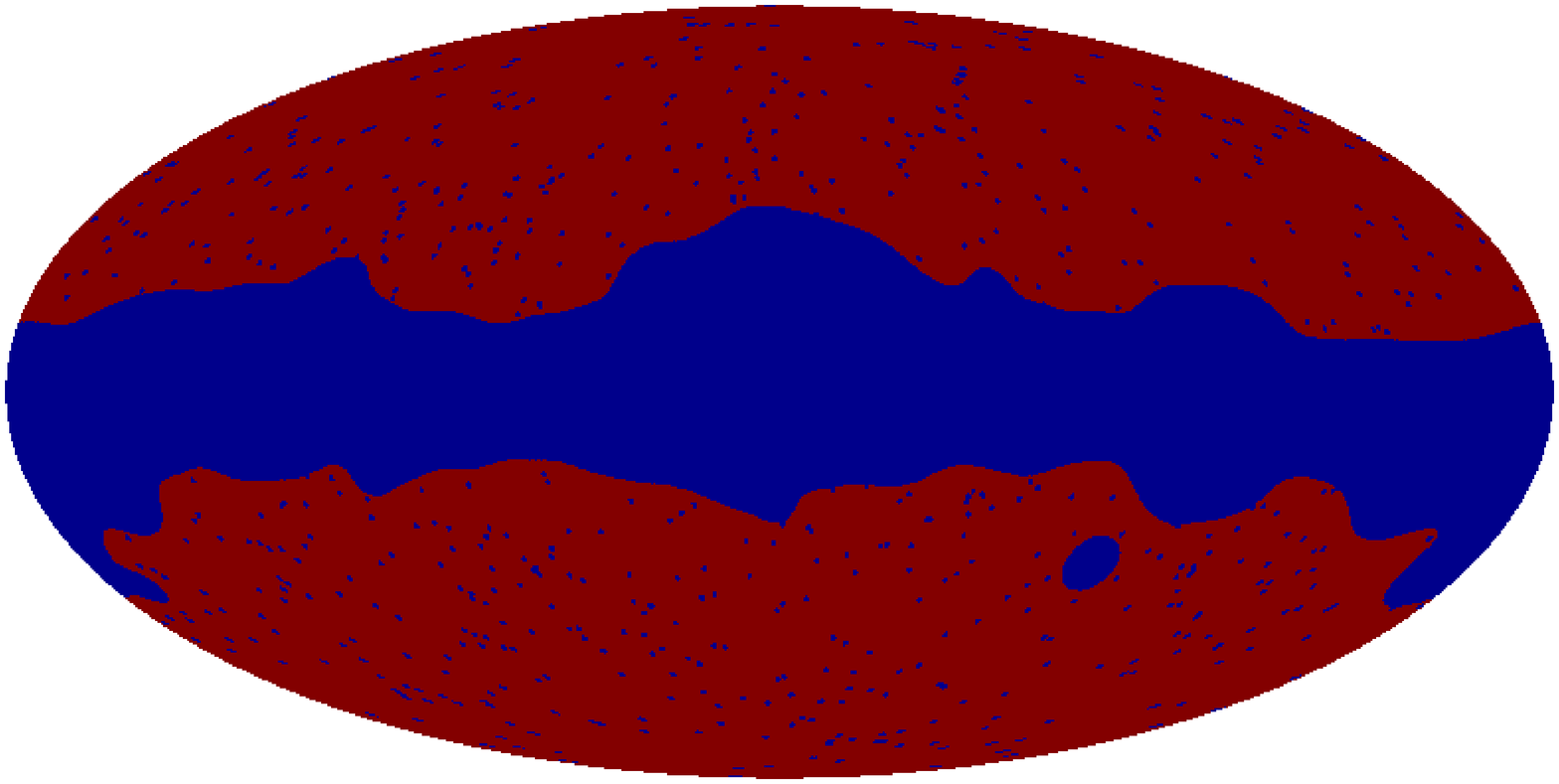}

\caption{The various masks used for computing the $f_{\rm NL}$ in Table~\ref{QVW_fnl_mask} . From left to right, and top to bottom we show Kp12,  Kp2, Kp0, and Kp0+. The point source exclusion regions are identical to those in the WMAP. Enlarging the point source exclusions does not change our results appreciably.}
\label{mask}
\end{figure}

To quantify the level at which foreground contamination might be expected to
affect our results, we perform a null test (see last row of Table~\ref{QVW_fnl_mask}). We apply our analysis to simulated
Gaussian CMB skies, expecting to measure $f_{\rm NL}$ consistent with 0. We 
model the expected diffuse synchrotron, dust and free-free emission
the same way the WMAP team did to produce their foreground cleaned maps. We
generate simulated WMAP Q, V and W maps including this foreground model, Gaussian
CMB and WMAP noise, and we found negligible $f_{\rm NL}$ ($-6\pm 4$ standard
error in the mean for 70 simulations) with $\ell_{max}=750$ and Kp0 mask. The
same analysis run on simulated Q+V+W data gives an $f_{\rm NL} \sim -13\pm 5$. $f_{\rm NL}$ becomes increasingly negative for smaller masks, following the
same pattern seen in the real data, and is negligible for the Kp0+ mask.

Note that in the cases that we expect to be affected by foregrounds
(e.g. small masks), $f_{\rm NL}$ is biased
negatively both in the real data and in simulations. It is therefore plausible that any bias due to residual foreground contamination 
would falsely {\it reduce} the significance of our detection. We assign a
systematic error of $\pm^6_0$ to diffuse foreground effects in V+W, kp0 and $\ell_{max}=750$.

To get an additional handle on foregrounds, we have also analyzed the WMAP foreground-reduced maps. Irrespective of the mask used we always find that foreground subtraction increases the $f_{NL}$ estimate. However, since we cannot guarantee that cleaning the foregrounds does not oversubtract foregrounds we conservatively quote the results from the raw maps.

For both the Kp0 and the extended masks, increasing $\ell_{max}$ increases the
central value of $f_{\rm NL}$. This effect may also be seen in Fig.~1. This change
is somewhat larger than what we measure in Monte Carlo simulations, but
attaching a significance to this observation is complicated by its {\it a
posteriori} nature.
In any case, foregrounds are likely not responsible since most models of the high latitude
galactic foreground emission predict more rapid decay of spatial structure in
these foregrounds compared to the CMB on the relevant scales.
Note that the same arguments also apply to a spinning dust component.

Our added statistical power relies on information at relatively small angular
scales, where unresolved point source emission may contaminate the map. There are several
reasons why it is unlikely to be responsible for our results. First, jointly
fitting for both the primordial and the point source bispectrum contributions
has shown that neglecting the point source contribution does not bias $f_{\rm NL}$
on the relevant range of scales
~\citep{KS2001,nong_wmap,SerraCooray08}. Second, since bright point sources are highly clustered,
one could hypothesize that there may be a contribution due to clustered point sources that
are not explicitly masked by the Kp0 mask but that are near the point source
exclusion regions. A re-analysis of the data using a mask with larger point
source exclusion regions showed no significant shift. Galactic point sources 
with a gradient in galactic latitude cannot be responsible because our estimate
is insensitive to extending the mask. 

All studies of the dominant secondary anisotropies conclude that they are
negligible for the analysis of the WMAP data for $l_{max}<800$ \citep{KS2001,SerraCooray08}. The largest expected 
biases arise from the ISW-lensing and SZ-lensing contributions.
These
partially cancel due to their opposite signs and are expected to contribute a
net bias of $\sim2$, and a much smaller effect on the error bar. 
Masking the Cold Spot found by~\cite{Cruz_etal_2005}
we find the central value of  $f_{\rm NL}$ increases to ($\sim 94$), but we ignore
this enhancement of our signal as an {\it a posteriori} effect.

Finally we study the dependence of $f_{\rm NL}$ on cosmological parameters. We find
that within the $2\sigma$ allowed range of cosmological parameters,
the central value and the error bar varies by than $10\%$ with the largest effect due to variations in $n_s$. Setting  $n_s=0.95$ reduces all kp0 estimates by $5-20\%$ while simultaneously reducing the variance by a similar amount. For $n_s=0.95$, $l_{max}=750$ has a significantly smaller variance than those for lower $l_{max}$ and gives $f_{\rm NL}= 83.5\pm 27$, increasing the statistical significance of
the detection to more than 3$\sigma$. 

We conclude that the WMAP 3 year data contains evidence that allows us to reject
the null-hypothesis of primordial Gaussianity at the 99.5\% significance level. Including our systematic error estimates, our result differs from $f_{NL}=0$ at the 2.5$\sigma$ level.
If our result holds up
under scrutiny and the statistical weight of future data, it will have profound
implications on our understanding of the physics of the early Universe. As it
stands, the data disfavors canonical single field slow-roll inflation. 

In addition to repeating our analysis on future data, further tests on currently available data using different higher order moments and additional bispectrum configurations may provide additional clues. The implications of our result for other probes of the cosmological
density field, such as the mass function and large scale structure data
should also be considered. This detection demonstrates the promise of targeted searches for primordial non-Gaussianity as a probe of the early Universe.

\bigskip
 
\acknowledgments We would like to thank E.~Komatsu for his help and 
advice during the project. The referees' questions helped us sharpen our tests
and arguments. We acknowledge using CMBFAST \citep{5} and HEALPix \citep{Healpix}. This  work  was  partially  supported by 
NCSA under TG-MCA04T015, by  NSF AST O5-07676 and NASA JPL subcontract 1236748,
the Alexander von Humboldt foundation, and
by the U.~of Illinois. We used the Teragrid Cluster
(www.teragrid.org).  We acknowledge the hospitality of MPA Garching 
where part of this work was done.


\end{document}